# Resonant diffusion of normal alkanes in zeolites: Effect of the zeolite structure and alkane molecule vibrations

[Roumen Tsekov](#) and P.G. Smirniotis

Chemical Engineering Department, University of Cincinnati, Cincinnati, Ohio 45221

Diffusion of normal alkanes in one-dimensional zeolites is theoretically studied on the basis of the stochastic equation formalism. The calculated diffusion coefficient accounts for the vibrations of the diffusing molecule and zeolite framework, molecule-zeolite interaction, and specific structure of the zeolite. It is shown that when the interaction potential is predominantly determined by the zeolite pore structure, the diffusion coefficient varies periodically with the number of carbon atoms of the alkane molecule, a phenomenon called resonant diffusion. A criterion for observable resonance is obtained from the balance between the interaction potentials of the molecule due to the atomic and pore structures of the zeolite. It shows that the diffusion is not resonant in zeolites without pore structure, such as ZSM-12. Moreover, even in zeolites with developed pore structure no resonant dependence of the diffusion constant can be detected if the pore structure energy barriers are not at least three times higher than the atomic structure energy barriers. The role of the alkane molecule vibrations is examined as well and a surprising effect of suppression of the diffusion in comparison with the case of a rigid molecule is observed. This effect is explained with the balance between the static and dynamic interaction of the molecule and zeolite.

Zeolites attract scientific attention because of their remarkable catalytic and sorption properties. They are regular porous materials with pore size in the order of several Angstroms. Because of their structure, zeolites can exhibit molecular shape selectivity to certain molecules, which is very important for many applications of practical interest[1]. An important factor for the efficiency of each process utilizing zeolites is the rate of intracrystalline mass transfer of species. For instance, chemical reactions are catalyzed in zeolites at some active centers and, therefore, are frequently diffusion controlled since the reactants have to reach these active centers first.

The methods to measure the mobility of molecules in zeolites are naturally divided in two groups[2]. The first group consists of measurements of some macroscopic integral response of the system. One can mention here, for instance, the classical absorption/desorption experiments. On this method the diffusion constant is calculated by comparing the experimental absorption/desorption curves with exact solutions of the unsteady state diffusion equation for the particular shape of zeolite grains employed[3]. The experimental curves, however, are influenced not by the diffusion process only. In fact, there are at least three consequent processes, namely, transfer of molecules from the gas phase to the zeolite, adsorption/desorption of the molecules on the zeolite grain surface, and diffusion in the zeolite crystal. If the activation energy of the second process is larger than that of the intracrystalline diffusion, the rate limiting stage will be the adsorption on (desorption from) the gas/crystal interface. In this case no reliable information about the diffusion in zeolites can be determined from the adsorption curves[4].

The second group consists of microscopic methods. Nuclear magnetic resonance and neutron scattering measurements[2] allow the determination of the diffusion coefficient in zeolites at equilibrium, i.e. without concentration gradients. However, because of the short observation time, the diffusion coefficients obtained could differ from those measured by the macroscopic methods. The latter is an average along the whole zeolite crystal and coincides at low loading with the tracer diffusion coefficient $D$ given by the Einstein formula $<X^2>=2Dt$ at large times. There are several time-scales in the Brownian motion through zeolites. At short times an alkane molecule travels a distance smaller than the period of the long-wavelength pore structure. In this case the diffusion cannot be resonant since it is governed by the atomic structure of the zeolite only. At medium time scale one can measure the diffusion coefficient $D$. At very large times other effects in real crystals occur such as scattering of the molecules from the walls, molecule-molecule impacts, etc. which certainly reflects in the value of the diffusion constant measured by macroscopic methods. Hence, to measure the macroscopic value of the diffusion coefficient microscopically, a minimal time of observation is required the order of which can be estimated by the ratio of the square lattice parameter of the zeolite and the diffusion constant. For large molecules it could be in the order of seconds. This remark is also essential for molecular dynamics (MD) simulations, which are usually restricted by computer capability to the range of nanoseconds[5]. Hence, it is not surprising that MD calculations agree with the results from spectroscopic measurements and are far from macroscopic ones[6]. This is also the reason for the applicability of MD simulations to short alkanes only, which diffuse relatively fast and thus the necessary observation time is small.

An interesting aspect of the diffusion of normal alkanes through zeolites is the resonant diffusion[7]. This phenomenon occurs only in ordered structures. Owing to the non-uniformity of the host media, the interaction potential of the diffusing molecule could depend non-monotonously on its geometric parameters. In this manner, it is possible that the energy surface of a long molecule exhibits lower barriers as compared to those of shorter molecules[8]. This will correspond to faster motion of the longer molecule, which is in contrast to the case in homogeneous media. Gorring[9] has reported experimental results for the diffusion of normal alkanes in T-zeolite where the diffusion constants exhibited a periodic dependence on the number of carbon atoms of the paraffin chain; a minimum at $C_8$ and a maximum at $C_{12}$. Obviously, an octane molecule does not fit well to the potential in T-zeolite, while a dodecane molecule is on an equipotential surface in the zeolite. The problem is that other researchers[10] fail to observe experimentally this kind of periodicity of the diffusion constant and even to reproduce[4] the results of Gorring.

On the other hand, recent MD simulations in silicalite[11] demonstrated the existence of periodic dependence of the diffusion coefficient on the number of carbon atoms of the diffusing normal alkanes. In fact, they exhibited resonance only in one of the components of the diffusion tensor. At present, the MD simulations represent computer experiments and their discrepancy with actual experiments indicate either wrong interpretation of the experimental results or unrealistic computational models. We are inclined to believe the MD simulations since the recent experiments by a new membrane technique have confirmed the resonant diffusion in silicalite[12]. To clarify the problem we have tried to propose an intermediate step by using an analytical approach. The present paper extends further some previous works[8,13] which involve a rigid diffusing molecule. We take into account the effects of alkane molecule vibrations and a more realistic potential of interaction with the zeolite. We demonstrate that the existence of sequence of expansions and apertures in the zeolite channels (see Figure 1) is necessary but not sufficient to observe resonant diffusion. If the energy barriers related to this zeolite pore structure are not sufficiently higher than the barriers induced by its atomic structure, the diffusion constant peaks are weak and practically undistinguishable. In addition, we show that the Brownian motion of a rigid molecule is faster than the motion of a vibrating one. This unexpected effect is due to the balance of static and dynamic interactions between the molecule and zeolite and was also observed in the diffusion of vibrating dimers on solid surfaces[14].

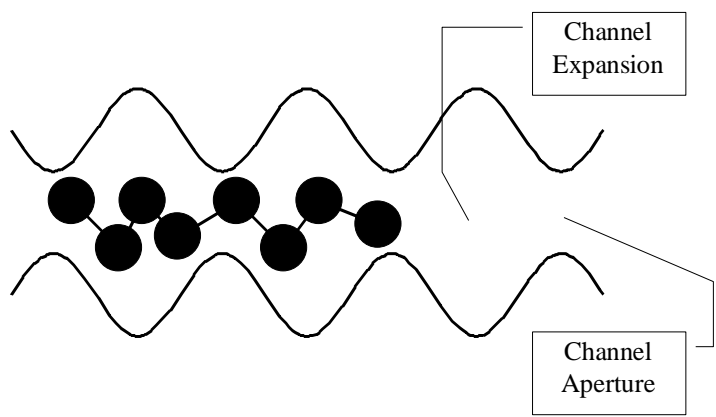

**Fig. 1** Scheme of a channel in the LTL zeolite hosting a normal octane molecule.

The present theory considers the diffusion of a single normal alkane molecule in zeolites and the tracer diffusion constant $D$ (called also self-diffusivity) is calculated. Hence, there will be no collective effects, which are inevitable in the experiments at finite loading of the sorbent. The motion of alkanes in zeolites is quite complicated. From the viewpoint of an alkane molecule, the zeolite is equivalent to a net of channels with different cross-sections. At least two processes are relevant to the motion of molecules in the zeolite network. The first is the Brownian motion of the molecule along a channel and the second is the turning to crossing channels. The mechanics of the turning, which occurs in 2D and 3D zeolite networks, is relatively complex to describe quantitatively. Alternatively, 1D zeolites such as ZSM-12 and Linde-type-L (LTL) zeolites are associated with motion only in one direction. The shape of a LTL channel[15] is presented in Figure 1 while ZSM-12 possesses channels of constant cross sectional area. The Brownian motion in these zeolites is one-dimensional and is confined in the frames of the occupied channel. We will restrict our consideration to this simpler case.

A self-consistent theoretical treatment of the Brownian motion of a molecule inside the zeolite crystal involves the description of the motion of the zeolite and molecule atoms interacting with each other. Since the hydrocarbon molecules and zeolite atoms are sufficiently heavy, the quantum effects will be neglected in our treatment. Hence, the classical mechanics definition of the problem is to specify the Hamiltonian function of the system. Within the scope of the harmonic approximation, the normal mode Hamiltonian function of the molecule-zeolite system is given by

$$H = \frac{P^2}{2M} + \frac{1}{2}(p^2 + q\Lambda^2 q) + \Phi(X,q) \qquad (1)$$

where $M$ is the mass of the molecule, $X$ and $P$ are the coordinate and the momentum of the molecule mass center along the zeolite $x$-axis coinciding with the occupied channel, $q$ and $p$ are the vectors of normal modes and associate momenta of the vibrations, respectively, $\Lambda^2$ is the matrix of the normal mode square frequencies of non-interacting molecule and zeolite, and $\Phi$ is the interaction potential. Equation (1) does not consider the possibility of torsional isomerization of the alkane molecule, which is justified by the high activation barrier of this process[6]. Since the amplitude of the vibrations is small, the potential $\Phi$ can be expanded in a series of $q$ to obtain

$$\Phi = \Phi_0 + (\partial_q \Phi)_0 q + \frac{1}{2} q (\partial_q \partial_q \Phi)_0 q \qquad (2)$$

The term $\Phi_0$ is the interaction potential of a rigid molecule and the static zeolite lattice. Note that in the present theory the rigorous quadratic term appears in eq 2. If it is strong enough, the interaction between the molecule and zeolite can cause either local deformations of the zeolite structure close to the molecule or substantial modulation of the molecule spectrum of vibrations. By introducing eq 2 into eq 1, the Hamiltonian function of the system acquires the form

$$H = \frac{P^2}{2M} + \frac{1}{2}(p^2 + q\Omega^2 q) + \Phi_0 + (\partial_q \Phi)_0 q \qquad (3)$$

where $\Omega^2 = \Lambda^2 + (\partial_q \partial_q \Phi)_0$ is the matrix of the spring constants in the coupled system. We will consider further that the local force constants do not depend substantially on the position of the molecule in the zeolite, i.e. $\Omega$ is a weak function of $X$.

According to the Hamiltonian function from eq 3 the equations of motion take the form

$$\ddot{q} + \Omega^2 q = -(\partial_q \Phi)_0 \tag{4}$$

$$M\ddot{X} = -\partial_X \Phi_0 - (\partial_X \partial_q \Phi)_0 q \tag{5}$$

Equation 4 describes the system vibrations under the action of an external force due to the molecule-zeolite interaction, while eq 5 is for the motion of the molecule mass center. Solving eq 4 and substituting the result for $q$ in eq 5 the latter changes to

$$M\ddot{X} + \int_0^t ds\, G\dot{X} + \partial_X U = F \tag{6}$$

Equation 6 is a particular example of the so-called generalized Langevin equation. The integral on the left-hand side represents the friction force and the memory function is given by

$$G = (\partial_X \partial_q \Phi)_0(t) \frac{\cos[\Omega(t-s)]}{\Omega^2} (\partial_q \partial_X \Phi)_0(s) \tag{7}$$

The effective potential of interaction between the molecule and zeolite is equal to

$$U = \Phi_0 - (\partial_q \Phi)_0 \frac{1}{2\Omega^2} (\partial_q \Phi)_0 \tag{8}$$

while the random Langevin force has the form

$$F = -(\partial_X \partial_q \Phi)_0 [\frac{\cos(\Omega t)}{\Omega^2} (\partial_q H)^0 + \frac{\sin(\Omega t)}{\Omega} (\partial_p H)^0] \tag{9}$$

where the superscript $^0$ denotes time zero. The stochastic nature of $F$ originates from unknown initial normal modes and pulses of the molecule and zeolite vibrations. Since an isothermal system is considered, the initial distribution of the vibrations should obey the Gibbs canonical law. Hence, according to eq 9 the average value of the Langevin force $F$ is zero, while its autocorrelation function is proportional to the memory function, $<F(t)F(s)> = kTG$. The last expression is known as the fluctuation-dissipation theorem, which is responsible for the constant value of the temperature $T$ in the system.

For the further analysis is essential to simplify the memory kernel expression. Since the potential $\Phi$ differs from zero only in the near environment of the molecule, eq 7 can be written in the form

$$G = (\partial_X \partial_q \Phi)_0 (\partial_q \partial_X \Phi)_0 \int_0^\infty \frac{\cos[\omega(t-s)]}{\omega^2} g_X(\omega) d\omega \tag{10}$$

while the effective potential from eq 8 is given by

$$U = \Phi_0 - (\partial_q \Phi)_0^2 \int_0^\infty \frac{g_X(\omega)}{2\omega^2} d\omega \tag{11}$$

The local density $g_X$ of normal modes with frequency $\omega$ refers to a small part of the coupled molecule-zeolite system containing the molecule and several layers of neighbor zeolite atoms. Apparently, the two important quantities above are directly related to the frequency spectrum of vibrations. However, $g_X$ is not equal to the sum of the spectral densities of the separate molecule and zeolite because of their interaction. Moreover, it represents the density of all vibrations, not limited to those, active in the IR-spectroscopy. Hence, the knowledge of the diffusing molecule and zeolite IR-spectra is not sufficient to model $g_X$ satisfactorily.

Usually, the characteristic time of memory is very short and for this reason it is unessential for the diffusion constant value calculated by the Einstein relation $D = <X\dot{X}>$ at infinite time. Hence, the slow motion of the molecule cannot follow the rapid vibrations of the environment and the generalized Langevin equation can be simplified to an ordinary one

$$M\ddot{X} + B\dot{X} + \partial_X U = F \tag{12}$$

where the friction coefficient is a time integral over the memory function and is given by the expression

$$B = (\partial_X \partial_q \Phi)_0^2 \int_0^\infty \frac{\sin(\omega t)}{\omega^3} g_X(\omega) d\omega = (\partial_X \partial_q \Phi)_0^2 \lim_{\omega \to 0} \frac{\pi g_X(\omega)}{2\omega^2} \tag{13}$$

The second more simplified form follows from the long time limit and the mathematical relation $\lim_{t \to \infty} \sin(\omega t)/\omega = \pi\delta(\omega)$. In this case the Langevin force $F$ reduces to a multiplicative white noise.

Since our theory is for the long times we expect also to be fulfilled the necessary condition for the overdamped limit of the molecule motion. Hence, one can derive from eq 12 the following form of the Smoluchowski equation

$$\dot{\rho} = \partial_x(\frac{\rho}{B}\partial_x U + \frac{kT}{B}\partial_x \rho) \tag{14}$$

which governs the evolution of the probability density $\rho$ to find the Brownian particle at a given point $x$ at time $t$. It is difficult to obtain a closed-form analytical solution of this equation, but we will develop here an efficient approximate method.

A formal solution of eq 14 in the region $x \geq 0$ is

$$\rho = \exp(-\beta U)\int_x^\infty dy \beta B \exp(\beta U)\int_y^\infty dz \dot{\rho} \tag{15}$$

where $\beta = (kT)^{-1}$ with $k$ being the Boltzmann constant. The probability density is a normalized quantity and thus by integration eq 15 can be transformed into

$$\frac{1}{2} = \int_0^\infty dx \exp(-\beta U)\int_x^\infty dy \beta B \exp(\beta U)\int_y^\infty dz \dot{\rho} \tag{16}$$

It should be noted that the potential $U$ and the friction coefficient $B$ are functions of $x$, which fluctuate around their constant values along the channel axis. If one is interested in calculating the diffusion constant for infinite time diffusion during which the molecule has visited all the points in the zeolite channel, the functions of $U$ and $B$ in eq 16 can be well estimated by their space-averaged values. Applying this procedure twice and integrating two times by parts, eq 16 simplifies to

$$1 = \overline{\exp(-\beta U)} \int_0^\infty dx^2 \beta B \exp(\beta U)\int_x^\infty dz \dot{\rho} = \overline{\exp(-\beta U)} \ \overline{\beta B \exp(\beta U)} <X\dot{X}> \tag{17}$$

where the bar indicates spatial average. Using now the Einstein definition $D = <X\dot{X}>$ for the diffusion coefficient, one gets the formula

$$D = [\ \overline{\beta B \exp(\beta U)} \ \overline{\exp(-\beta U)}\ ]^{-1} \tag{18}$$

This expression allows calculation of the diffusion coefficient in any modulated structure. It is also applicable to the case when the molecule changes channels; the time spent in perpendicular channels can be modeled by proper potential wells at the channel cross sections. Equation 18 was derived rigorously by Festa and d'Agliano[16] for the case of a periodic potential $U$. At low temperature, the first term in eq 18 is proportional to the maximum of the potential, while the second one is more sensitive to its minimum. Hence, the diffusion coefficient should depend exponentially on their difference divided by temperature, which is the Arrhenius law. Note that $D$ is more sensitive to the value of $B$ on the top of the potential barriers than in the potential wells. This is essential because some zeolites possess $U$-maxima in their channel expansions while others exhibit $U$-maxima in their channel apertures where the friction is higher.

To calculate the diffusion coefficient in a LTL zeolite is necessary to provide expressions for the potential $U$ and the friction coefficient $B$. A realistic model of the alkane molecule is that of united "atoms", i.e. one can consider it as a chain of methylene group "atoms". If $\phi(x)$ is the potential energy of a methylene group in the zeolite then the potential of static interaction can be presented as

$$\Phi_0 = \sum_{j=1}^{N} \phi(X + x_j) \tag{19}$$

where $x_j = (2j - N - 1)a/2$ is the average position of the $j$-"atom" refered to the mass center $X$, $N$ is the number of carbon atoms in the chain, and $a$ is the distance between two methylene groups along the molecule axis. Since the interaction potential decreases rapidly with increasing distance between the zeolite atoms and alkane molecule, $\Phi$ differs essentially from zero in the very neighborhood of the molecule and the vibrational terms can be estimated as

$$(\partial_q \Phi)_0^2 \approx \frac{1}{\mu} \sum_{j=1}^{N} [\partial_X \phi(X + x_j)]^2 \qquad (\partial_X \partial_q \Phi)_0^2 \approx \frac{1}{\mu} \sum_{j=1}^{N} [\partial_X^2 \phi(X + x_j)]^2 \tag{20}$$

where $\mu$ is an effective mass of the vibrating atoms.

Finally, one needs an expression for the spectral density $g_X$. As mentioned above, the exact form of the local spectral density is unclear. A favorable circumstance is that it contributes integrally to eq 11 and hence its exact form is not necessary for our analysis. The Debye model, $g(\omega \leq \omega_D) = 3\omega^2 / \omega_D^3$ and $g(\omega > \omega_D) = 0$ with $\omega_D$ being the Debye frequency, is a crude description of the spectral density of vibrations in solids. Previous considerations[8] show, however, that this one-parametric model is too rough for description of the diffusion in zeolites. In the present case we expect even more sufficient deviations of $g_X$ from the Debye model due to the vibrations of the molecule. Therefore, we simply introduce the following two characteristic frequencies

$$\frac{1}{\omega_0^3} = \lim_{\omega \to 0} \frac{g_X(\omega)}{3\omega^2} \qquad \frac{1}{\omega_\infty^2} = \int_0^\infty \frac{g_X(\omega)}{3\omega^2} d\omega \qquad (21)$$

Note that in the Debye approximation these two frequencies coincide with the Debye one. We will consider further $\omega_0$ and $\omega_\infty$ as two independent variables which require at least a two-parametric model for the spectral density $g_X$.

Introducing eqs 19-21 into eq 11 and eq 13 one obtains more detailed expressions for the interaction potential and the friction coefficient

$$U = \sum_{j=1}^{N} \phi(X + x_j) - \frac{3}{2\mu\omega_\infty^2} \sum_{j=1}^{N} [\partial_X \phi(X + x_j)]^2 \qquad (22)$$

$$B = \frac{3\pi}{2\mu\omega_0^3} \sum_{j=1}^{N} [\partial_X^2 \phi(X + x_j)]^2 \qquad (23)$$

Hence, the last model needed is one for the methylene group-zeolite interaction potential $\phi$. The methylene group potential surface can be traced approximately by the diffusion of methane in the zeolite. In general, there are two kinds of periodicity in the zeolite crystal. The short length periodicity is due to the discrete atomic structure. Occasionally, the lengths of the Si-O and C-C bonds are very close and, for this reason, one can accept the length $a$ of the C-C bond as the period of the zeolite atomic structure. The longer length periodicity is due to the pore structure units, such as expansions and apertures. Hence, a reasonable model for the interaction potential of a methylene group in LTL is

$$\phi(x) = A\cos(2\pi x/na) + C\cos(2\pi x/a) \qquad (24)$$

Here a simple cosine dependence on the position is assumed and the pore structure period is expressed in $a$-units. The exact potential of a methylene group in LTL calculated my molecular simulations could differ from expression (24). The latter can be considered as the leading terms in the Fourier series expansion of the exact potential. Hence, there is no problem to estimates the parameters $A$ and $C$ if the exact potential is known but to calculate the latter is not a trivial work. Usually, it is easy to calculate the potential minima, which are important for description of the adsorption in zeolites[16]. The number $n$ is usually larger than 5 and if the coefficients $A$ and $C$ are commensurable, it is clear that the leading term in the derivatives of $\phi$ is the last one. Hence, in the calculations of the friction coefficient $B$ from eq 23 and the vibrational term of the effective potential $U$ from eq 22 one can neglect the first term in eq 24. This is not surprising since both the friction and vibrations are atomic scale phenomena and naturally more sensitive to the atomic structure of the zeolite. Thus, the following analytical expressions can be obtained

$$U = A \frac{\sin(\pi N/n)}{\sin(\pi/n)} \cos(2\pi\varphi) + CN \cos(2\pi n\varphi) - \frac{6\pi^2 C^2 N}{\mu \omega_\infty^2 a^2} \sin^2(2\pi n\varphi) \qquad (25)$$

$$B = \frac{24\pi^5 C^2 N}{\mu \omega_0^3 a^4} \cos^2(2\pi n\varphi) \qquad (26)$$

with $\varphi = X/na$.

Let us consider first the expression of the friction coefficient. Obviously it is proportional to the strength $C$ of the interaction between the molecule and zeolite atoms and inversely proportional to the frequency of vibrations (in the limit $\omega_0 \to \infty$ all the vibrations in the system become frozen and, of course, there is no friction). Since the alkane motion is one-dimensional and the Si-O and C-C bond lengths are practically the same all the methylene groups undergo the same friction. Hence, the friction coefficient depends linearly on their number as prescribed by eq 26. The situation with the potential is quite different. Because the C-C and Si-O bond lengths are approximately equal, there are two terms being linear functions of $N$. The last one accounts for the vibrations and can be meaningful only if the frequency $\omega_\infty$ is smaller than a certain value, $\mu \omega_\infty^2 a^2 \leq 6\pi^2 C$. The first term, however, depends periodically on the number of carbon atoms of the alkane. It is zero if $N$ is an integer times larger than $n$ and its absolute value is maximal if the ratio $N/n$ is a half number. This is a precondition for resonant diffusion.

If $n$ is an integer, the functions $B(\varphi)$ and $U(\varphi)$ are periodic with period of one. Hence, the spatial-averaged values can be taken for this period and the diffusion constant will be given by

$$D = [\int_0^1 d\varphi \ \beta B \exp(\beta U) \int_0^1 d\varphi \exp(-\beta U)]^{-1} \qquad (27)$$

Unfortunately, the analytical integration of eq 27 is impossible. In order to investigate the effect of various parameters of our theory, we will further perform numerical calculations. Some of the parameters needed will be fixed, namely, $a$ = 1.25 Å, $\mu$ = 3x10$^{-26}$ kg, $C$ = 2x10$^{-21}$ J and $\omega_0$ = 5x10$^{12}$ Hz. Since the period of repeating expansions in a LTL channel is 7.47 Å the corresponding perio-dicity number *n* is about 6. Our variable parameters are the number of carbon atoms *N*, temperature, pore structure potential constant *A* and frequency $\omega_\infty$. In order to simplify the presentation, we introduce the following dimensionless parameters $\alpha = A/C$, $\gamma = \beta C$ and $\nu = \omega_\infty/\omega_0$ (note that at the fixed values of the parameters above, the important range of $\omega_\infty$ is $\nu \leq 3$). In the range of the temperatures considered we expect the diffusion constant to follow the Arrhenius law and we calculate also the activation energy and pre-exponential factor by the relations

$$\varepsilon_a = -\partial_\beta \ln(D) \qquad D_0 = D\exp(\beta\varepsilon_a) \qquad (28)$$

For the purpose of convenience the activation energy will be presented per mol, $E_a = N_A \varepsilon_a$, where $N_A$ is the Avogadro number.

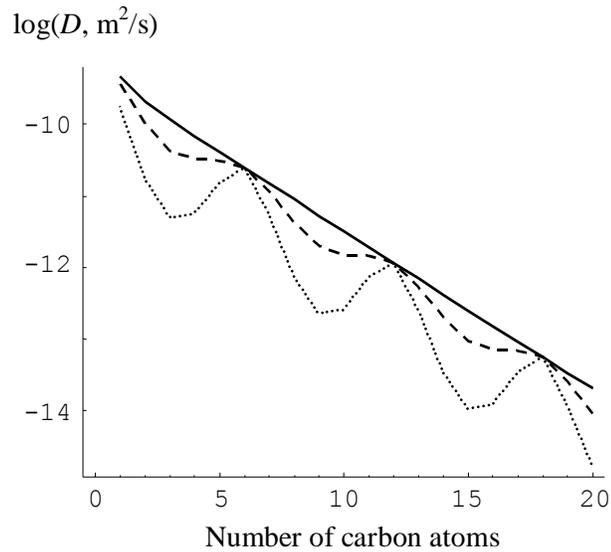

**Fig. 2** Dependence of the diffusion coefficient *D* on the number *N* of carbon atoms of the alkane molecule at $\gamma = 0.25$, $\nu = 10$, $\alpha = 0$ (solid line), $\alpha = 3$ (dashed line) and $\alpha = 6$ (dotted line).

Figure 2 shows the dependence of the diffusion coefficient on the carbon number for different values of $\alpha$. One can observe that at relatively large $\alpha$ the dependence of *D* on the number of carbon atoms of the alkane is periodic with the maxima for $C_6$, $C_{12}$, and $C_{18}$ and the minima for $C_3$, $C_9$, and $C_{15}$. When the value of $\alpha$ becomes smaller, the minima and maxima become less pronounced and finally disappear at $\alpha = 0$. This means that in zeolites without expansions and apertures like ZSM-12 ($\alpha = 0$) there will be no resonant dependence of the diffusion coefficient *D*. Moreover, our calculations show that the resonant effects are essential if $\alpha \geq 3$. Hence, even in zeolites with well developed pore structure no resonant dependence of *D* can be measured if the difference of adsorption energy between the expansions and apertures is not sufficiently higher than the atomic structure energy barriers. Due to the equality of the Si-O and C-C bond lengths, the effect of the zeolite atomic structure is similar to that of a homogeneous media and causes decrease of $D$ with increasing molecule size.

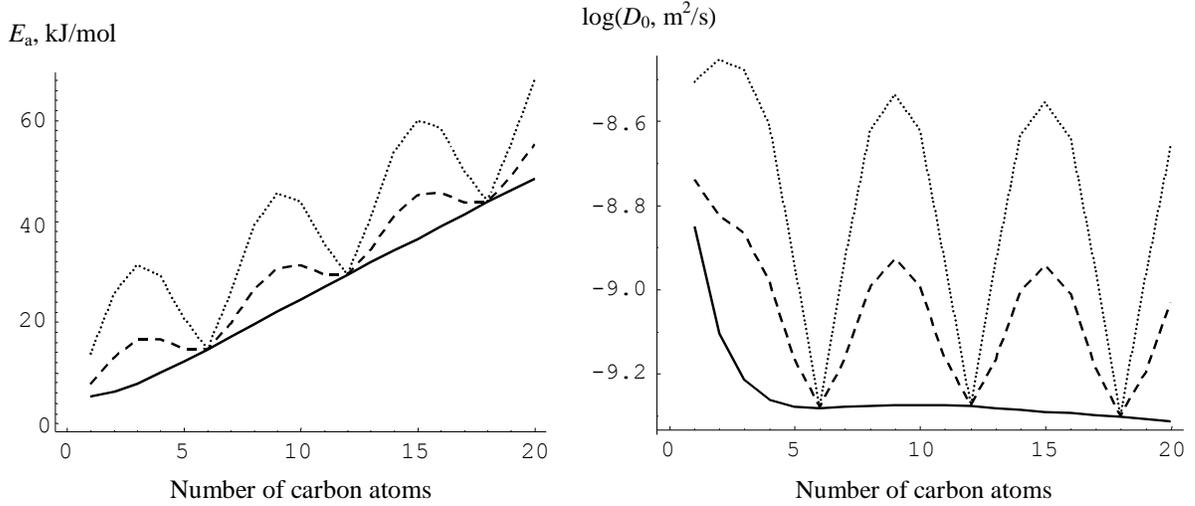

**Figs. 3** and **4** Dependence of the activation energy $E_a$ and the pre-exponential factor $D_0$ on the number $N$ of carbon atoms of the alkane molecule at $\gamma = 0.25$, $\nu = 10$, $\alpha = 0$ (solid line), $\alpha = 3$ (dashed line) and $\alpha = 6$ (dotted line).

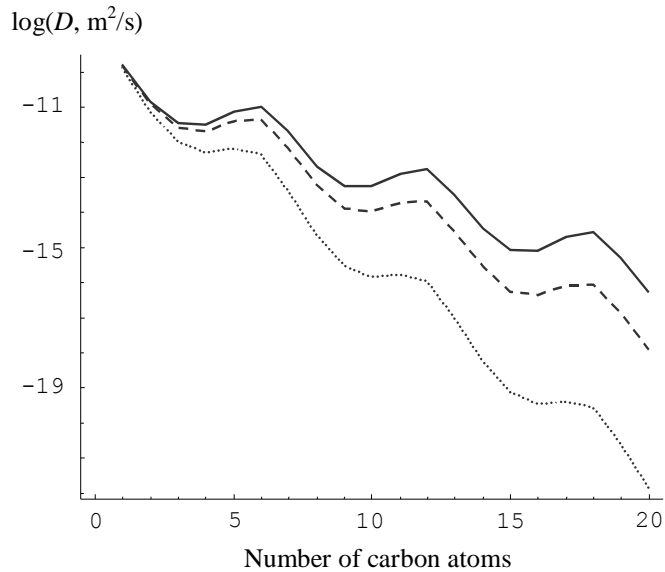

**Fig. 5** Dependence of the diffusion coefficient $D$ on the number $N$ of carbon atoms of the alkane molecule at $\gamma = 0.25$, $\alpha = 6$, $\nu = 2.5$ (solid line), $\nu = 2$ (dashed line) and $\nu = 1.5$ (dotted line).

The activation energy and pre-exponential factor corresponding to the diffusion constant from Figure 2 are presented in Figure 3 and Figure 4, respectively. Both of them are also periodic functions of N for $\alpha$ different from zero. The activation energy increases by increasing carbon number while the pre-exponential factor just fluctuates around a constant value. The latter fact implies proportionality between $D_0$ and $E_a$, since this is the only way to cancel the effect of the linear dependence of the friction coefficient B on N. Hence, the maxima of the activation energy actually correspond to maxima of $D_0$ as presented in the figures. Note that the effects of the pre-exponential factor and activation energy on the diffusion constant vary in the opposite directions. Thus, the increase of $D_0$ with increasing the activation energy is a compensation effect. At temperature of 300 °C, however, the activation energy is more important than the pre-exponential factor and the diffusion constant exhibits maxima at the $E_a$-minima.

The results on the previous figures are calculated at $\nu = 10$, i.e. the vibrational term of the potential from eq 25 contributes negligibly to the diffusion. The zeolite lattice is usually considered at rest in the MD simulations in order to save computational time. This approximation is correct only if the molecule is much more flexible than the zeolite, since in this case the values of $\omega_0$ and $\omega_\infty$ are those of the molecule. Hence, the diffusion coefficient is no longer sensitive to the zeolite vibrations, which are very rapid to be followed by the molecule. To study the effect of the molecule vibrations the dependence of the diffusion constant on N at different $\nu$ is presented in Figure 5. As seen the decrease of $\nu$ decreases the diffusion coefficient and the effect is more pronounced for larger molecules. The corresponding activation energy and pre-exponential factor are presented on Figure 6 and Figure 7, respectively. The effect of the frequency $\omega_\infty$ is unexpected because the conventional understanding is that flexible molecules are diffusing faster than rigid ones. By decreasing $\nu$ we in fact make the molecule more flexible (the case of a rigid molecule and rigid zeolite corresponds to infinite $\omega_\infty$). This means that the vibrations of the molecule lead to an increase of the activation energy in comparison with the rigid analog (the obvious increase of the friction is excluded here since we kept $\omega_0$ to be constant). The fact that the flexible molecules move slower than the rigid ones is not trivial and deserves a special attention. A plausible explanation can be seen by analysing eq 25. For rigid molecules the second term of eq 25 dominates and the activation energy amounts to $\varepsilon_a = 2CN$. For very flexible molecules the dominant term is the third one, and the activation energy in this case is equal to $\varepsilon_a = 6\pi^2 C^2 N / \mu \omega_\infty^2 a^2$. However, since the last term dominates if $\mu \omega_\infty^2 a^2 \leq 6\pi^2 C$, it is clear now that the activation energy in the second case could be larger than that of the rigid molecule.

Finally, Figure 8 presents the dependence of D on N at different temperatures. By increasing temperature the effect of the activation energy on the diffusion constant naturally decreases. After a certain temperature it no longer dominates the opposite effect of the pre-exponential factor and the diffusion constant becomes non-periodic with respect to the carbon number of the diffusing alkane. Such a decrease of the resonant effects by increasing temperature is also observed in MD simulations[9]. For the present system the critical temperature above is estimated to be about 1200 °C. In Figure 8 we took sufficiently large $\alpha$ constant to magnify the resonant diffusion and for this reason the corresponding critical temperature is high. In practice, however, it could be within the range of the experimental measurements.

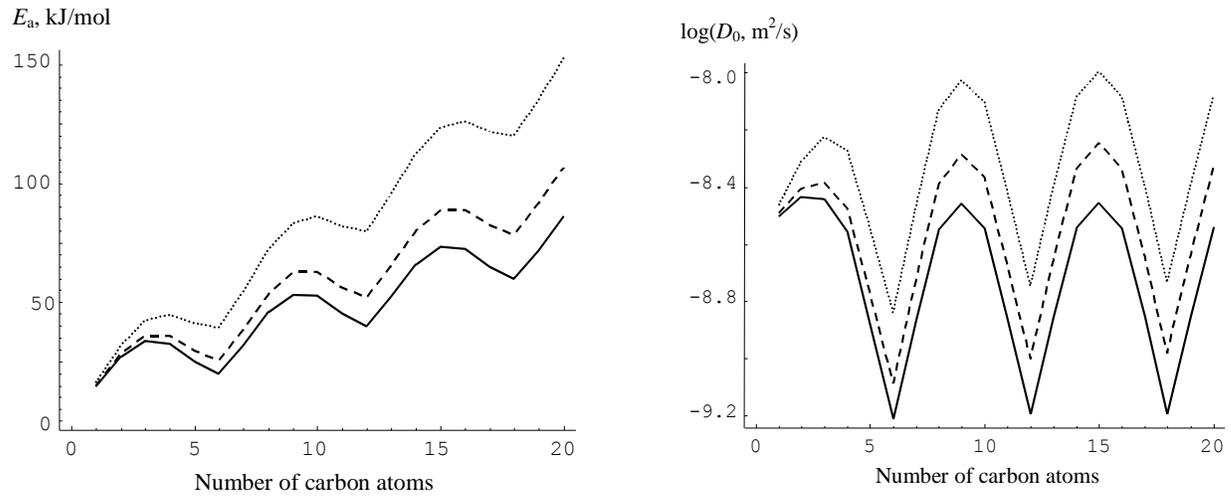

**Figs. 6** and **7** Dependence of the activation energy $E_a$ and the pre-exponential factor $D_0$ on the number $N$ of carbon atoms of the alkane molecule at $\gamma = 0.25$, $\alpha = 6$, $\nu = 2.5$ (solid line), $\nu = 2$ (dashed line) and $\nu = 1.5$ (dotted line).

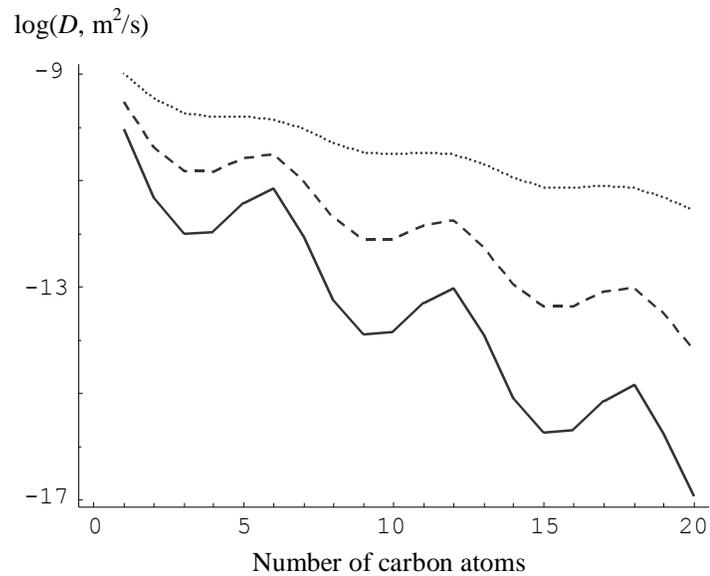

**Fig. 8** Dependence of the diffusion coefficient $D$ on the number $N$ of carbon atoms of the alkane molecule at $\alpha = 6$, $\nu = 3$, $\gamma = 0.30$ (solid line), $\gamma = 0.20$ (dashed line) and $\gamma = 0.10$ (dotted line).

The present study demonstrated the importance of the zeolite structure on the diffusion of species through zeolites. The resonant diffusion of normal alkanes in zeolites is due to the zeolite pore structure of rings, cages, cross sections, expansions, apertures, etc. However, if the energy barrier related to these units is not sufficiently higher than that due to the atomic structure, the resonant diffusion is weak and difficult to observe. This is likely the case in nature since the magnitude of the potential interaction between a methylene group and a zeolite[17] is of the order of 3 kJ/mol and the activation barrier for diffusion of methane[6] is 5 kJ/mol . Hence, a periodic dependence of the diffusion constant of alkanes on the number of their carbon atoms is an exception for peculiar zeolites, rather than a rule. As mentioned, the MD simulations[10] show that only one of the components of the diffusion tensor exhibits resonant diffusion in silicalite. The macroscopic value of $D$, however, depends on all the three components and it could be practically insensitive to the resonant component. If this is the case, no resonant diffusion will be detected in a macroscopic experiment.

The important effect of the alkane vibrations has also been demonstrated in the present paper. It can reflect on the decrease of the specific frequencies $\omega_0$ and $\omega_\infty$. The decrease of $\omega_0$ increases the friction constant and in this way the Brownian motion of the molecule is slowed. The decrease of $\omega_\infty$ increases mainly the activation energy (see Figure 6 and Figure 7) which also leads to a smaller diffusion constant. Hence, the alkane molecule vibrations decrease the molecule diffusivity in comparison with the rigid analog. Our model, however, is for one-dimensional motion of the molecule along a straight channel. In a more complicated case involving turning to other channels the flexibility of the molecule is important for this process and certainly will accelerate the diffusion. Finally, an interesting effect elucidated in this paper is the compensation effect of the pre-exponential factor. It leads to suppression of the resonant diffusion, and thus for any zeolite there is a critical temperature below which the resonant diffusion exists. Hence, if the experiments are carried out at a temperature above the critical one, no resonant diffusion will be detected. It is clear that because of the quasi-phenomenological character of our approach no comparison with experiments is possible. The goal of the developed theory is to "read" the experimental results and to "translate" them in a microscopic level. Then the specific constants such as $A$ and $C$ extracted from experimental data can be compared with exact MD calculations.

1) Chen, N.Y.; Garwood, W.E.; Dwyer, F.G. *Shape-Selective Catalysis in Industrial Applications*; Dekker: New York, 1989.
2) Karger, J.; Ruthven, D.M. *Diffusion in Zeolites and Other Microporous Solids*; Wiley: New York, 1992.
3) Barrer, R.M. *Adv. Chem.* **1971**, *102*, 1.
4) Magalhaes, F.D.; Laurence, R.L.; Conner, W.C. *AIChE J.* **1996**, *42*, 68.
5) Demontis, P.; Suffritti, G.B. *Chem. Rev.* **1997**, *97*, 2845.
6) Maginn, E.J.; Bell, A.T.; Theodorou, D.N. *J. Phys. Chem.* **1996**, *100*, 7155.

7) Ruckenstein, E.; Lee, P.S. *Phys. Lett. A* **1976**, *56*, 423.
8) Tsekov, R.; Ruckenstein, E. *J. Chem. Phys.* **1994**, *100*, 1450, 3808.
9) Gorring, R.L. *J. Catal.* **1973**, *31*, 13.
10) Cavalcante Jr., C.L.; Eic, M.; Ruthven, D.M.; Occelli, M.L. *Zeolites* **1995**, *15*, 293.
11) Runnebaum, R.C.; Maginn E.J. *J. Phys. Chem. B* **1997**, *101*, 6394.
12) Talu, O.; Sun, M.S.; Shah, D.B. *AIChE J.* **1998**, *44*, 681.
13) Nitsche, J.M.; Wei, J. *AIChE J.* **1991**, *37*, 661.
14) Tsekov, R. *NATO ASI Ser. B Phys.* **1997**, *360*, 419.
15) Fyfe, C.A.; Gies, H.; Kokotailo, G.T.; Marler, B.; Cox, D.E. *J. Phys. Chem.* **1990**, *94*, 3718.
16) Festa, R.; d'Agliano, E.G. *Physica A* **1978**, *90*, 229.
17) Maginn, E.J.; Bell, A.T.; Theodorou, D.N. *J. Phys. Chem.* **1995**, *99*, 2057.